\documentclass{article}

\usepackage[preprint]{neurips_2025}

\usepackage[utf8]{inputenc} 
\usepackage[T1]{fontenc}    
\usepackage{hyperref}       
\usepackage{url}            
\usepackage{booktabs}       
\usepackage{amsfonts}       
\usepackage{nicefrac}       
\usepackage{microtype}      
\usepackage{xcolor}         
\usepackage{natbib}
\usepackage{algorithm}
\usepackage{algpseudocode}
\usepackage{amsmath}
\usepackage{dsfont}
\usepackage{wrapfig}
\usepackage{graphicx}
\usepackage[skip=1pt, font=small]{caption}
\usepackage{enumitem}

\title{Cross-device Zero-shot Label Transfer via Alignment of Time Series Foundation Model Embeddings}

\author{%
  Neal G. Ravindra\\
  Independent Researcher\\
  \texttt{ngravindra@gmail.com}\\
  \And
  Arijit Sehanobish\\
  Independent Researcher\\
  \texttt{arijit.sehanobish1@gmail.com}\\
}

\begin{document}

\maketitle



\begin{abstract}
 High-quality, medically validated labels exist for clinical actigraphy data but not for ubiquitous consumer wearables like the Apple Watch. Manually labeling wearables data is expensive and doesn't scale. This paper offers a novel framework that transfers valuable labels from a source domain (e.g., actigraphy) to a target domain (e.g., Apple Watch) without requiring paired data. Instead of working with raw time-series signals, we project both domains into a shared latent embedding space using time-series foundation models (TSFMs) and develop a new framework to align the cross-device representations. Our method, \emph{Adversarial Alignment of TSFM Embeddings} forces the distributions of source and target embeddings to align within this space, facilitating label transfer across device type.
\end{abstract}

\section{Introduction}

A significant challenge in digital health is that high-quality, validated labels often exist for data from clinical-grade devices but are not easily transferable to data from ubiquitous consumer wearables. For example, medically validated labels from actigraphy studies are rarely available for devices like the Apple Watch, limiting the reach of research findings \citep{ravindra2023deep}. Working with raw time-series signals is difficult, and transferring labels between misaligned domains at consumer scale is a major bottleneck to fully leveraging wearable devices' time-series for health-care.

\begin{wrapfigure}{r}{0.6\columnwidth}
  \vspace{-15pt}
  \centering
  \includegraphics[width=0.6\columnwidth]{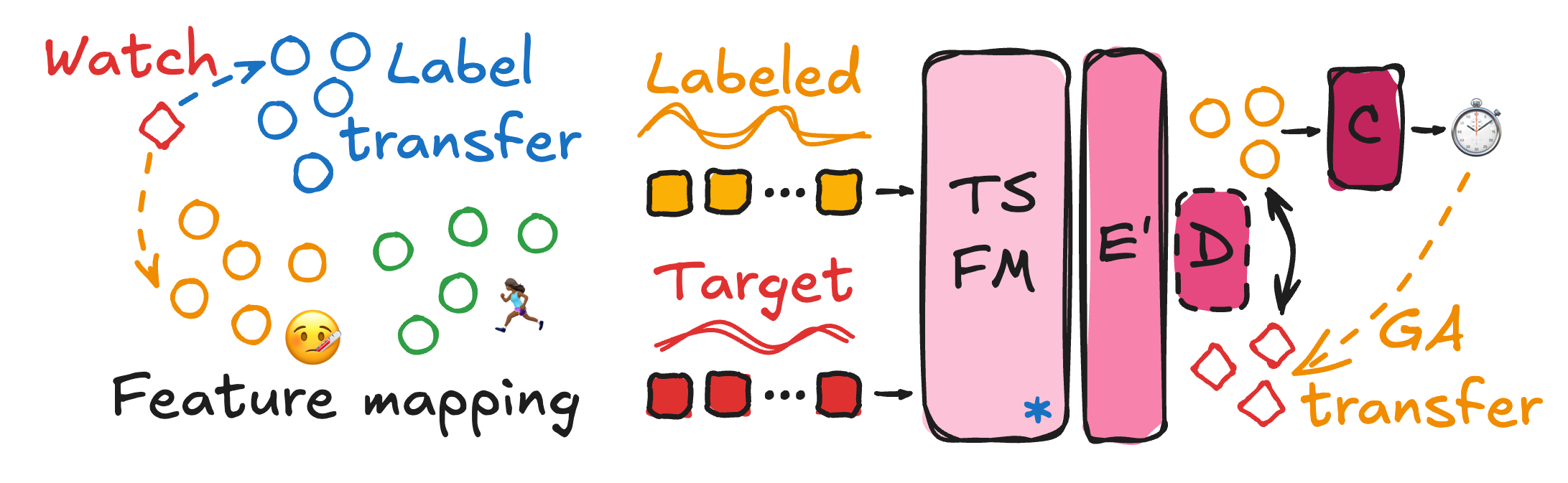}
  \caption{Aligning TSFM Embeddings for Label Transfer. Our framework uses a pre-trained TSFM to embed source (labeled) and target (watch, un-labeled) data. A trainable adapter ($E^{\prime}$) and a domain discriminator ($D$) align the embeddings, enabling a task classifier ($C$) to perform zero-shot transfer of the gestational age label (GA) to consumer devices.}
  \label{fig:overview}
  \vspace{-10pt}
\end{wrapfigure}

Recent work has shown that while the raw accelerometry data between devices can be comparable \citep{white2024comparison}, significant domain shifts persist, complicating direct model transfer. Currently, a surge of Time-Series Foundation Models (TSFMs) have demonstrated remarkable success in forecasting tasks \citep{vaswani2017attention, zhou2021informer, nie2023time, anas2024chronos}. These models excel at forecasting because they learn fundamental temporal patterns like trends and seasonality. The rise of powerful TSFMs has created a new class of rich, semantic embeddings for time-series data. However, it's an open question whether these embeddings are suitable for standard, off-the-shelf domain adaptation techniques to solve real-world problems. \textit{We hypothesize that these powerful representational capabilities are composable and transferable; a model that can forecast a time series must inherently understand its underlying structure, making its internal embeddings a rich source of features for other downstream tasks, such as the clinical regression task we address.} We investigate this by pairing a state-of-the-art TSFM (Chronos) with a classic domain adaptation method (adversarial alignment). Our work is an empirical validation showing that this combination is indeed effective for a challenging cross-device label transfer task in healthcare.

\par 

\textbf{Our contributions} are: (1) A novel framework for zero-shot label transfer between wearable devices using a frozen TSFM backbone and a lightweight, adversarially-trained adapter. (2) A principled simulator that generates consumer-grade data by iteratively degrading person-specific signatures from clinical-grade data, guided by a pre-trained patient identifier model. (3) A demonstration of successfully transferring a clinical regression model for Gestational Age prediction from a clinical to a consumer-grade domain, showcasing a clear performance rescue over a non-aligned baseline.

\textbf{Related works}: Our work is positioned at the intersection of: (1) time-series representation learning, (2) domain adaptation for wearables, and (3) healthcare applications. Self-supervised methods like TS2Vec have shown success in learning universal time-series representations from unlabeled data \citep{yue2022ts2vec}. Several TSFMs have demonstrated success in forecasting \citep{zhou2021informer, nie2023time} but not in regression and domain adaptation, the focus of our work \cite{potosnak2025investigating}. Models like N-BEATS explicitly learn to decompose signals into fundamental basis functions for trend and seasonality \citep{oreshkin2019n} but do not explore trends in wearables data. Currently, large-scale models have been pre-trained on massive datasets of unlabeled wearable data to learn general representations \citep{kiyasseh2023scaling,narayanswamy2024scaling}. Our work differs from these in two fundamental ways. First, while these approaches demonstrate the power of scale for learning from unlabeled data, our problem setting focuses on transferring scarce, high-cost, and medically-validated clinical labels to domains where they are absent. Second, rather than using a model pre-trained only on physiological signals, we leverage general-purpose TSFMs pre-trained on a vast and heterogeneous corpus of time series from many fields \citep{anas2024chronos}. 

\section{Adversarial Alignment of Foundation Model Embeddings}

\textbf{Problem set-up}: We have a high-quality dataset $\mathcal{D}_i = (X_S, Y_S)_i$ consisting of time-series data $X_i$, for example, raw, clinical-grade actigraphy accelerometry data over time, and labels associated with that raw time-series signal $Y_i$, for example, various clinical indications and outcomes.

\begin{wrapfigure}{r}{0.6\columnwidth} 
  \vspace{-15pt}
  \centering
  \includegraphics[width=0.55\columnwidth]{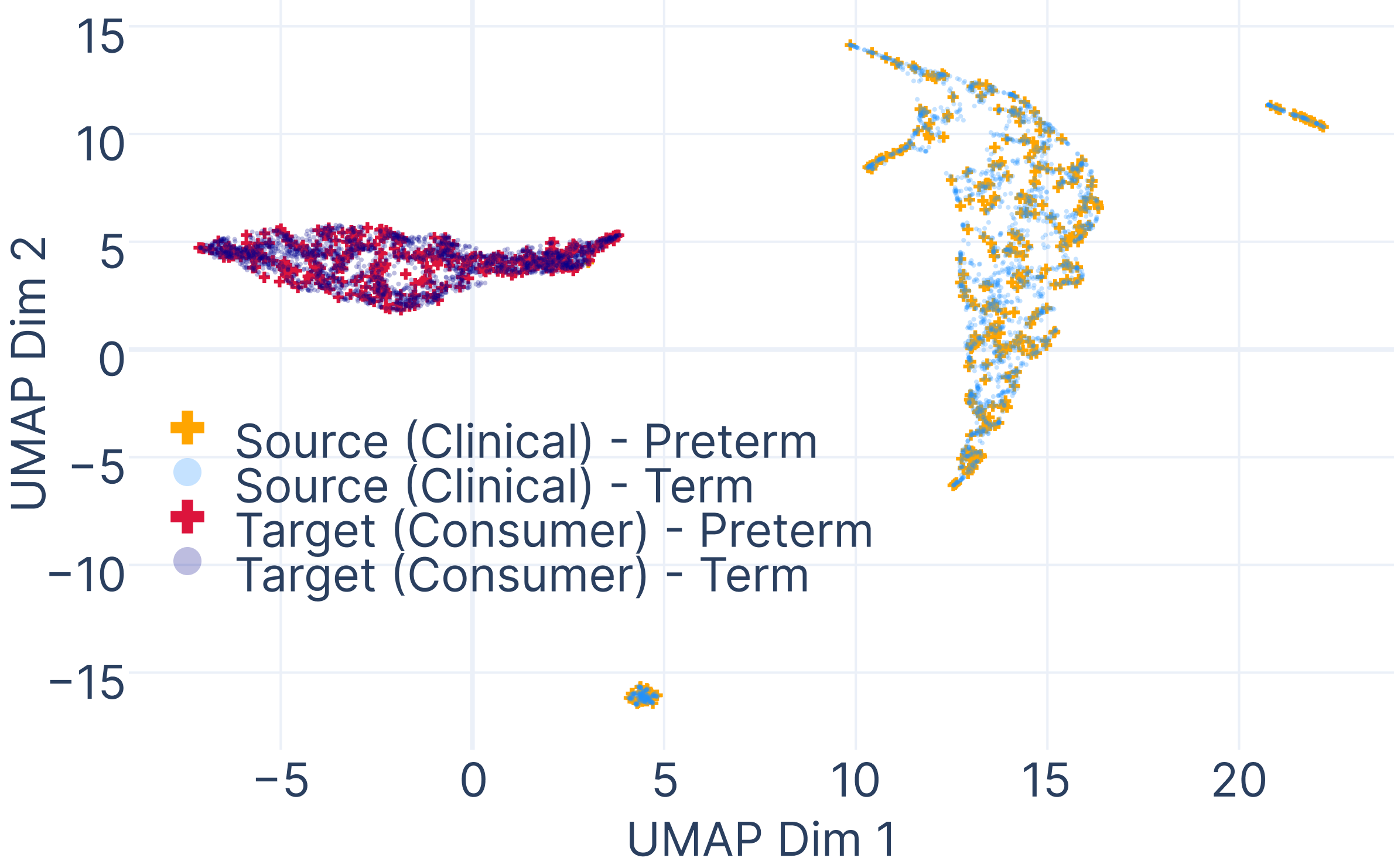}
  \caption{UMAP projection of embeddings from the baseline model (average output of TSFM, left). Points are colored by their domain, Source (Clinical) vs. Target (Consumer), and styled by the clinical outcome.}
  \label{fig:umap_shift}
  \vspace{-10pt}
\end{wrapfigure}


To transfer labels, the ideal scenario would be a paired dataset where each subject simultaneously wears both a clinical-grade actigraphy device and a consumer wearable (e.g., an Apple Watch). However, such datasets are rare, and we do not have consumer wearable data for the specific subjects in our clinical source cohort. Therefore, to bridge this gap without paired data, we develop a generator to simulate a realistic, unlabeled target domain. For each source sample $X_S$, our generator creates a corresponding consumer-grade representation, $X_T$, enabling investigation of the domain alignment process.

\subsection{Apple Watch Generator} \label{sec: apple watch generator}

To simulate lower-quality, consumer-grade time-series data, we propose a domain adaptation framework based on noise modeling, up-sampling, and signal processing. The procedure for generating an unlabeled target stream from source data is detailed in Algorithm~\ref{alg:data_adaptation}. Briefly, our simulator transforms a source patch $X_S$ into its consumer-grade counterpart $X_T$ by introducing two key effects: (i) reduced signal-to-noise ratio and (ii) obfuscation of patient-specific features. To achieve the latter, we train a patient identification model $C$ as an “anonymization scorer.” During simulation, the generator iteratively perturbs a source patch with noise until the scorer’s confidence in correctly identifying the patient falls below a fixed threshold, thereby emulating the feature degradation expected from proprietary consumer devices.

\begin{algorithm}[h!]
\caption{Iterative Source Anonymization and Domain Adaptation}
\label{alg:data_adaptation}
\begin{algorithmic}[1]
\Require Source patch $X_S$ from patient $ID_S$, pre-trained identifier $C$, anonymization threshold $\delta$, max iterations $N_{max}$, noise parameter $\sigma$.
\State \textbf{Initialize}: $X_{iter} \leftarrow X_S$
\State Calculate baseline identifiability score: $S_{base} \leftarrow \text{Score}(C, X_S, ID_S)$
\State Set target score threshold: $S_{thresh} \leftarrow S_{base} \cdot (1 - \delta)$
\Statex
\Statex \textbf{Anonymization Loop:}
\For{$i = 1 \to N_{max}$}
    \If{$\text{Score}(C, X_{iter}, ID_S) \le S_{thresh}$} \textbf{break} \EndIf
    \State Inject adaptive noise: $X_{iter} \leftarrow X_{iter} + \text{GaussianNoise}(0, \sigma \cdot (1 + i / N_{max}))$
\EndFor
\Statex
\Statex \textbf{Final Domain Simulation:}
\State Apply smoothing, magnitude rescaling, and random masking to $X_{iter}$ to get $X_T$.
\State \Return $X_T$
\end{algorithmic}
\end{algorithm}


\textbf{Implementation details}: For our scoring model, we train a patient identification classifier, $C$, using the embeddings from our TSFM feature extractor. We provide these embeddings, $E(X_S)$, to an AutoML pipeline (`autogluon`) to train a robust ensemble model that serves as the scoring function in Algorithm~\ref{alg:data_adaptation}~\citep{erickson2020autogluon}.

\subsection{Time-series Foundation Model Backbone} \label{method: feature extractor}
In our method, TSFMs serve as a feature extractor. Specifically, we adapt pre-trained models (e.g., time-series transformers such as Chronos) to define the feature mapping $E$. Since these models are trained on diverse time-series data, they capture general structural and statistical properties, making them suitable for extracting informative representations in our setting. Finally assume that we have two streams of data: (1) \textbf{Source domain}: high-quality, labeled data $(X_S, Y_S)$, and (2) \textbf{Target domain}: unlabeled data, $X_T$.
\begin{align}
    (z_S, z_T) &= (E(X_S), E(X_T)) \label{eqn:tsfm_encoder}
\end{align}

\textbf{Implementation details}: We first use a frozen, pre-trained Chronos TSFM (chronos-t5-large), $E$, to map both source ($X_S$) and target ($X_T$) patches into a rich embedding space. Our model, $E'$, consists of this frozen backbone plus a trainable adapter network. Following the parameter-efficient strategy proposed by \citet{houlsby2019parameter}, our adapter is a lightweight MLP with a bottleneck structure that projects the 1024-dimensional TSFM embedding down to our final 128-dimensional aligned space. The core of our method is an adversarial game between this adapter and a domain discriminator, $D$. Since paired clinical-grade and consumer (e.g., Apple Watch) wearable data is unavailable, we generate our target domain data, $X_T$, using a novel simulator. To mimic the characteristics of a lower-quality consumer device, our framework transforms a source signal patch $X_S$ by introducing several degradations. These include \textbf{smoothing} to simulate on-device processing, adaptive \textbf{noise injection} to lower the signal-to-noise ratio, and random \textbf{masking} of signal segments to represent periods of non-wear. The full procedure is detailed in Section \ref{sec: apple watch generator}.

\subsection{Adversarial Domain Alignment and Label Transfer}

\textbf{Adversarial Domain Alignment}: We introduce a simple domain discriminator, $D$, which takes as an input, one of the paired signal representations from Equation \ref{eqn:tsfm_encoder} and outputs a prediction as to the origin. $D$ has the objective of minimizing its classification as to whether the embedding came from source or target domain.
\begin{equation}
    D = E^{\prime}(z) \in \mathds{1}(\text{source}) \label{eqn: domain adaptation}
\end{equation}

The encoder $E$, equipped with learnable adapters $E^{\prime}$, is trained to produce embeddings that fool the discriminator. Its goal is to construct a domain-invariant embedding space in which source and target embeddings, $z_S^{\prime}$ and $z_T^{\prime}$, are indistinguishable while preserving semantic content.

\textbf{Adversarial Training Details}: We use a Mean Squared Error loss for the domain discriminator, following the LSGAN approach for improved stability \citep{mao2017least}. To ensure the discriminator remains a strong opponent, its learning rate is set to twice that of the adapter's optimizer. The classification and adversarial losses are balanced by a hyperparameter $\lambda$.

\textbf{Label Transfer}: In parallel with the adversarial game, we train a classifier head $C$ downstream of the adapter $E^{\prime}$, using \emph{only} the aligned source embeddings and their corresponding labels $(z_S', Y_S)$. Because the adversarial training forces the unlabeled target embeddings $z_T'$ into the same representational space as the source, the decision boundary learned by the classifier on the source data naturally generalizes to the target domain, enabling zero-shot label transfer at inference time.


\vspace{-1em}
\section{Experiments and Results}
We begin by presenting the experimental setup, followed by results and analysis.
\vspace{-1em}
\subsection{Experimental Setup}
To evaluate whether our method successfully transfers labels from the source to the adapted target domain, we design a framework for zero-shot label transfer on a clinical prediction task using clinical-grade wearable data and associated metadata.~\cite{ravindra2023deep} (see Fig.~\ref{fig:overview}).

\begin{wraptable}{r}{0.5\textwidth}
  \vspace{-10pt}
  \caption{Quantification of domain mixing before and after alignment.}
  \label{tab:mixing_results}
  \centering
  \begin{tabular}{lcc}
    \toprule
    \textbf{Model} & \textbf{Entropy $\uparrow$} & \textbf{ARI $\downarrow$} \\
    \midrule
    Baseline (Before) & 0.002 & 0.226 \\
    Our Method (After) & \textbf{0.957} & \textbf{0.005} \\
    \bottomrule
  \end{tabular}
  \vspace{-10pt}
\end{wraptable}

\textbf{Data \& Task}: We use an actigraphy dataset from a study of pregnant individuals with corresponding gestational age (GA) at delivery~\citep{ravindra2023deep}. We extract 7-day patches from the 1-minute activity count data. The task is to predict the final GA at delivery from an activity patch taken during pregnancy. We frame this as a regression task, implemented by classifying the GA into 38 weekly bins. Our target domain data ($X_T$) is generated using the simulator described in Section~\ref{sec: apple watch generator}.

\begin{wrapfigure}{l}{0.4\columnwidth}
  \vspace{-15pt}
  \centering
  \includegraphics[width=0.4\columnwidth]{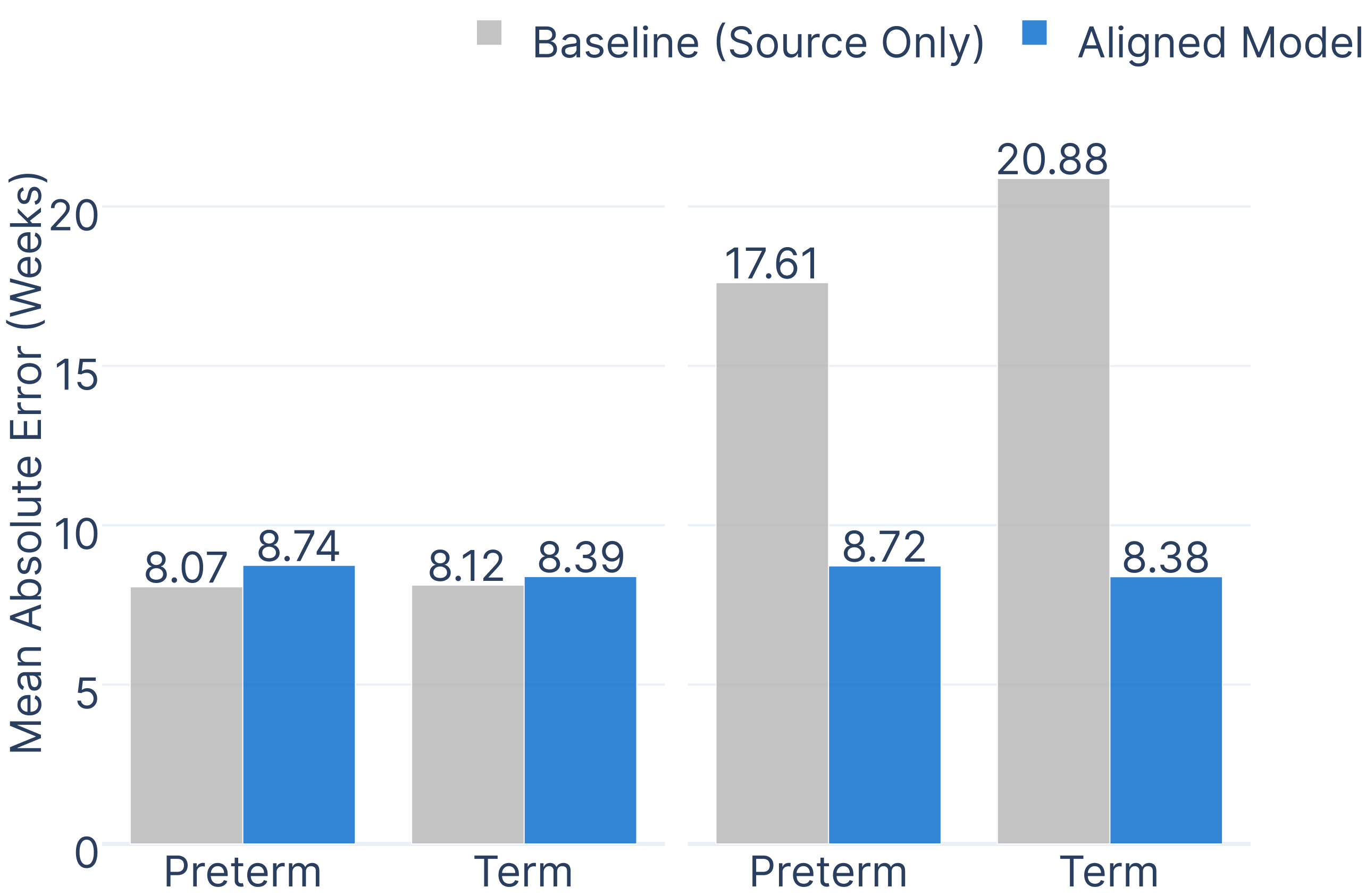}
  \caption{Domain Alignment Rescues Prediction Performance of Label Transfer. The plot compares Mean Absolute Error (MAE) for GA prediction on the source and target domains for the Baseline (Source Only) model versus our full Aligned model.}
  \label{fig:main_result}
  \vspace{-10pt}
\end{wrapfigure}

\textbf{Results.} We compare our method against a "Source Only" baseline, where the adversarial alignment is disabled. Figure \ref{fig:main_result} shows the key results. The baseline model performs reasonably on the source domain but suffers a catastrophic performance collapse on the target domain, with the Mean Absolute Error (MAE) more than doubling. Our adversarial alignment method successfully mitigates this, maintaining a low MAE on the target domain that is nearly identical to its source domain performance. This demonstrates a successful transfer of predictive capability. The UMAP visualizations in Figure \ref{fig:umap_shift} highlight the initially separated domains becoming well-mixed after alignment. Table \ref{tab:mixing_results} quantifies that, post-alignment, the domains are well-mixed, showing a dramatic increase in the domain mixing entropy score and a near-zero separation score (ARI) for our aligned model.

\section{Discussion and Limitations}

Our work demonstrates a practical solution to a major bottleneck in digital health: leveraging sparsely labeled clinical datasets for large-scale analysis of consumer wearable data. By operating on the rich, semantic latent spaces produced by TSFMs, our method moves beyond classic domain adaptation of raw signals. This creates a pathway to translate findings from controlled studies to real-world populations. Future work will need to prospectively validate these findings.


\bibliographystyle{plainnat}
\bibliography{references}

\end{document}